\documentclass[reprint,superscriptaddress,preprintnumbers,amsmath,amssymb,aps,prd,tightenlines,longbibliography,nofootinbib]{revtex4-2}

\usepackage{graphicx}
\usepackage{xcolor}
\usepackage[sort&compress]{natbib}
\usepackage{amsmath,amssymb,bm,bbm,slashed,subdepth,amsfonts}
\usepackage{xr-hyper}
\usepackage[colorlinks=true
,urlcolor=blue
,anchorcolor=blue
,citecolor=blue
,filecolor=blue
,linkcolor=red
,menucolor=blue
,linktocpage=true
,pdfproducer=medialab
,pdfa=true
]{hyperref}
\usepackage{cleveref}
\usepackage{enumerate}
\usepackage{epsfig, subfigure}
\usepackage{setspace}
\usepackage{booktabs, tabularx}
\usepackage{units}
\usepackage{placeins}
\usepackage{multirow}
\usepackage{mathtools}
\usepackage[normalem]{ulem}

\newcommand{\dyc}[1]{\textcolor{black}{#1}}

\begin{document}

\title{Beyond the Starobinsky model after ACT}

\author{Min Gi Park}
\email{mgpark@yonsei.ac.kr}
\affiliation{Department of Physics and IPAP, Yonsei University, Seoul 03722, Republic of Korea}

\author{Dhong Yeon Cheong}
\email{dycheong@uchicago.edu}
\affiliation{Department of Physics and IPAP, Yonsei University, Seoul 03722, Republic of Korea}
\affiliation{Enrico Fermi Institute, The University of Chicago, 5640 S Ellis Ave, Chicago, IL 60637, USA}
\affiliation{Kavli Institute for Cosmological Physics, The University of Chicago, 5640 S Ellis Ave, Chicago, IL 60637, USA}

\author{Seong Chan Park}
\email{sc.park@yonsei.ac.kr}
\affiliation{Department of Physics and IPAP, Yonsei University, Seoul 03722, Republic of Korea}
\affiliation{School of Physics, Korea Institute for Advanced Study, Seoul 02455, Korea}


\date{\today}

\begin{abstract}
We revisit higher order corrections to the Starobinsky inflationary model using the most recent P-ACT-LB-BK18 data, which exhibits a mild but definite tension with the predictions of the original model. 
Our results demonstrate how even small {higher order} deformations of the Ricci scalar (e.g. $R^3, R^4,\cdots$) can bring the model into better agreement with current data and impose nontrivial constraints on the post-inflationary dynamics.
\end{abstract}


\maketitle

\section{\label{sec:intro}Introduction}

Cosmic inflation provides a compelling explanation for the initial conditions of the Universe, accounting for its flatness, homogeneity, and the origin of primordial perturbations~\cite{Sato:1980yn, Guth:1980zm, Linde:1981mu}. Among the wide class of inflationary models (see e.g.,~\cite{Nojiri:2017ncd, Kallosh:2025ijd}), the Starobinsky model~\cite{Starobinsky:1980te}, based on a simple addition of $R^2$ term on top of the Einstein-Hilbert action 
\begin{align}
    S=\frac{1}{2}\int d^4 x \sqrt{-g} ~\left[R + (\beta/2) R^2\right], 
\end{align} 
stands out due to its remarkable agreement with CMB observations, particularly those from the \textit{Planck} observation~\cite{Planck:2018jri}.
Here we set the unit of the Planck mass $M_P=1/\sqrt{8\pi G} = 1$. 

However, recent improvements in observational precision on e.g. the spectral index $n_s$, especially with the recent data from the Atacama Cosmology Telescope (ACT)~\cite{ACT:2025tim}, have revealed a mild but significant ($\gtrsim 2\sigma$) tension  with the predictions of the original Starobinsky model with 60 efoldings:
{
\begin{align}
n_s \simeq \begin{cases}
0.975 \pm 0.005 &\text{(P-ACT-LB-BK18)}\\
0.965 &\text{(Starobinsky, $R^2$)}
\end{cases}
\end{align}
Not much improvement has been achieved by ACT regarding the tensor-to-scalar ratio $r\lesssim 0.038$ ($95\%$, P-ACT-LB-BK18).
}
A number of theoretical efforts have recently been undertaken to explain the data~\cite{Aoki:2025wld, Gialamas:2025kef, Drees:2025ngb, Kim:2025dyi, Zharov:2025zjg,  Liu:2025qca, He:2025bli, Choi:2025qot, Yogesh:2025wak, Cacciapaglia:2025xqd, Heidarian:2025drk, Addazi:2025qra, Haque:2025uri, Gialamas:2025ofz, Odintsov:2025bmp,Modak:2025bjv, Ferreira:2025lrd, Wang:2025dbj, Zharov:2025evb}.

Previous work of some of the authors~\cite{Cheong:2020rao}   has pointed out that an extension involving higher order curvature terms, 
\begin{align}
f(R) = R + (\beta/2) R^2 + \sum_{n=1}^\infty \frac{c_{n+2}}{n+2} R^{n+2} 
\end{align}
can lead to observable deviations in inflationary predictions, such as the spectral index $n_s$  and the tensor-to-scalar ratio $r$. The Starobinsky limit is recovered when we take $c_3=c_4=c_5 =\cdots =0$.

In this work, we revisit the same model allowing only the non-vanishing cubic term $c_3\equiv \gamma \neq 0$, $c_{n> 3}=0$ in light of the more recent ACT data.~\footnote{When $c_{n<N}=0$,  the first non-vanishing next order term $c_{N+1}\neq 0$ will be the most important contribution, in general.} 
We reanalyze its inflationary predictions using a perturbative expansion in the deformation parameter 
\begin{align}
\delta \equiv \frac{c_3}{c_2^2} \ll 1
\label{eq: delta}
\end{align}
where we take $\beta=c_2$ for brevity of notation. 
In particular, we study the implications of this deformation for the post-inflationary reheating phase. Using the standard parametrization of reheating via the effective equation-of-state parameter  $w_{\rm eff}$  and the reheating temperature  $T_{\mathrm{re}}$~\cite{Cook:2015vqa, Cheong:2021kyc}, we derive the relations among  $\delta$,  $N_e$, and $T_{\mathrm{re}}$, and identify the regions in parameter space that remain consistent with the observed scalar amplitude and spectral index. Our results show that the inclusion of the cubic term not only improves the fit to CMB data but also leads to nontrivial constraints on the reheating dynamics, providing a consistent picture of early-universe evolution beyond the Starobinsky model.

The paper is organized as follows. Section~\ref{sec:R3inflation} introduces the cubic $R^3$ deformation and derives the inflationary parameters. Section~\ref{sec:Trehdeltarange} analyzes the relation between the reheating temperature and inflationary predictions.
Conclusions are given in Section~\ref{sec:conclusions}.


\section{Starobinsky+ \texorpdfstring{$R^3$}{R³} Inflation}
\label{sec:R3inflation}

In this section, we extend the analysis by incorporating higher-order corrections to the Ricci scalar. Treating these terms perturbatively, we construct the effective potential and derive the resulting inflationary predictions.

\subsection{Set-up}

We first consider a general $f(R)$ gravity and its equivalent description with a scalar field $\phi$. 
The action is given as
\begin{align}
    S &= \frac{1}{2} \int d^4x \sqrt{-g} f(R) \\
      &\rightarrow \int d^4x \sqrt{-g} \left[ \frac{1}{2} \Omega^2(\phi) R - V(\phi) \right]
\end{align}
where $\Omega^2(\phi)=f'(\phi)$ and $V(\phi)=\frac{1}{2}(f'(\phi)\phi-f(\phi))$. The resultant action includes the non-minimal coupling term~\cite{Futamase:1987ua}, which can provide a flat potential in Einstein frame $V_E= V/\Omega^4\to constant$ at a high scale $\phi \to \infty$~\cite{Park:2008hz}. 

Explicitly, the corresponding Einstein-frame action is obtained via a Weyl rescaling $g_{E\mu\nu} = \Omega^2 g_{\mu\nu}$, yielding
\begin{align}
    S_E = \int d^4x \sqrt{-g_E} \left[ \frac{1}{2} R_E - \frac{1}{2} g_E^{\mu\nu} \partial_\mu s \partial_\nu s - V_E(s) \right],
\end{align}
where the canonical scalar field $s$ and the potential $V_E(s)$ are given by
\begin{align}
    s(\phi) &= \sqrt{\frac{3}{2}} \ln\left(\Omega(\phi)^2\right) =\sqrt{\frac{3}{2}}\ln\left(f'(\phi)\right),\\
    V_E(s) &= \frac{V(\phi(s))}{\Omega(\phi(s))^4}=\frac{\phi f'(\phi)-f(\phi)}{2f'(\phi)^2}.
\end{align}

\begin{figure}[t]
    \centering
    \includegraphics[width=0.95\linewidth]{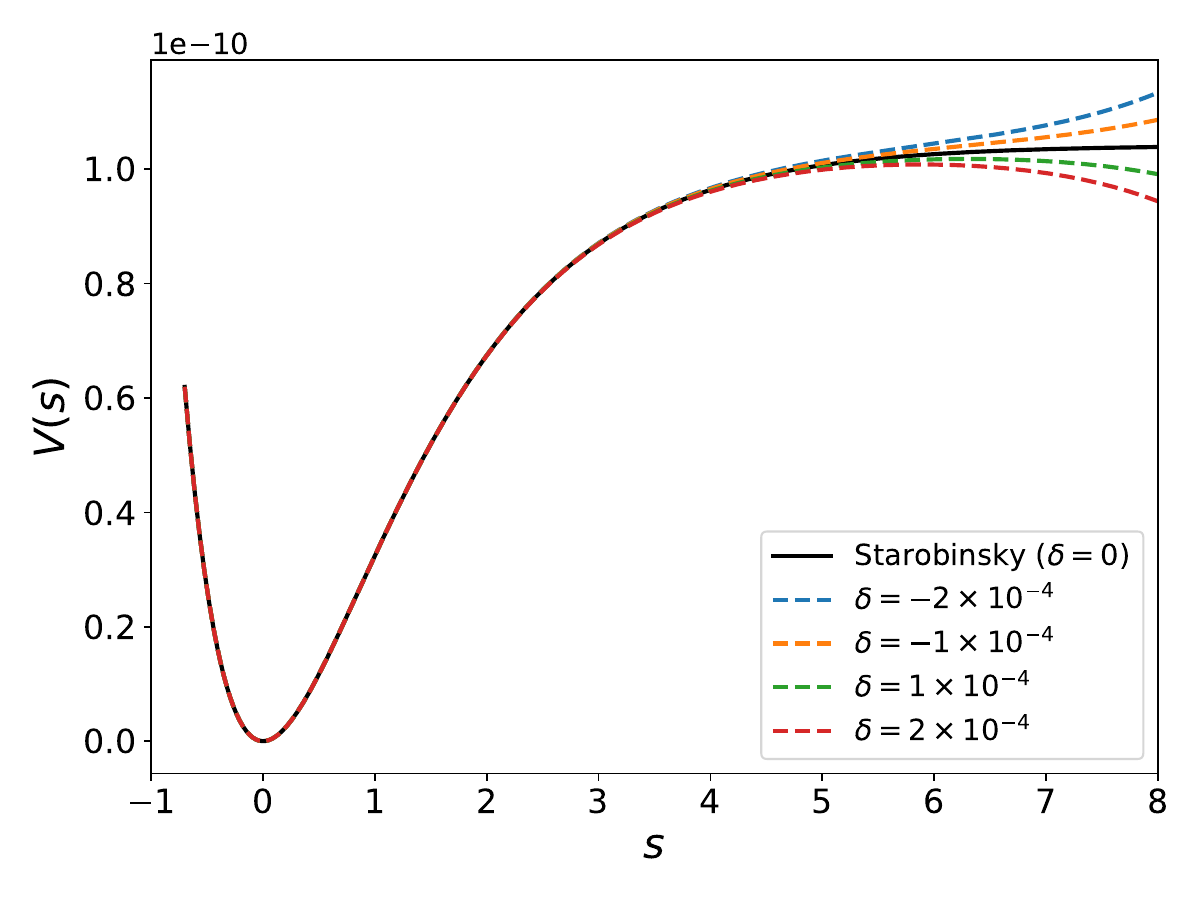}
    \caption{{The Einstein–frame potential $V_E(s)$. 
The black solid line represents the original Starobinsky model ($\delta = 0$), 
while the colored dashed lines correspond to the  $R^3$ term included potential as in Eq.~\eqref{eq:Einstein_V_exact}.}}
    \label{fig:potential}
\end{figure}

Now for our case, we have $f(\phi)=\phi+(\beta/2)\phi^2+(\gamma/3)\phi^3$ taking the cubic term into account. The size of the higher-order terms is required to be suppressed, as discussed in ~\cite{Jinno:2019und}. For the given situation, correspondingly, we get $s(\phi)=\sqrt{\frac{3}{2}}\ln(1+\beta\phi+\gamma\phi^2)$ or  $\phi(s)=\frac{\beta}{2\gamma}\left(\sqrt{1+4\delta(\sigma(s)-1)}-1\right)$ with 
    $\sigma(s)=e^{\sqrt{\frac{2}{3}}s}$. 
The potential in the Einstein frame is
\begin{align}
    V_E&=\frac{\beta\phi(s)^2\left(1+\frac{4\gamma}{3\beta}\phi(s)\right)}{4\left(1+\beta\phi(s)\left(1+\frac{\gamma}{\beta}\phi(s)\right)\right)^2}\\
    \label{eq:Einstein_V_exact}
    &\approx V_0(s)\left[1-\frac{2}{3} \delta\left(\sigma(s)-1\right)+\cdots\right],
\end{align}
where $\delta\equiv\gamma/\beta^2$ as defined in Eq.~\eqref{eq: delta} and $V_0=\frac{1}{4\beta}(1-e^{-\sqrt{\frac{2}{3}}s})^2$ is the potential of  the original potential of the Starobinsky model or equivalently the Higgs inflation~\cite{Bezrukov:2007ep}, especially, the critical Higgs inflation~ \cite{Hamada:2014iga,Hamada:2014wna}. (see \cite{Cheong:2021vdb} for a review.) The consistency with quantum gravity is discussed in e.g.~\cite{Cheong:2018udx, Lee:2021cor}. 
{ Fig.~\ref{fig:potential} shows the Einstein-frame potential $V_E$ for various values of $\delta$. For $\delta>0$, the correction lowers \dyc{while approaching a concave potential}, whereas for $\delta<0$ it \dyc{increases and gives a convex potential} in the large-$s$ region. These deformations modify the slow-roll parameters $\epsilon$ and $\eta$ and consequently shift the inflationary observables $n_s$ and $r$, which we elaborate more in the following sections.}

\subsection{Inflationary Parameters}
We treat the $\gamma$-dependent contribution as a perturbation, and expand the slow-roll parameters as
\begin{align}
    \epsilon &= \frac{1}{2} \left( \frac{V_E'}{V_E} \right)^2 = \epsilon_0 + \delta \Delta \epsilon, \\
    \eta &= \frac{V_E''}{V_E} = \eta_0 + \delta \Delta \eta,
\end{align}
where $\epsilon_0$, $\eta_0$ correspond to the standard Starobinsky model ($\gamma = 0$). Explictly, these quantities can be expressed as:
\begin{align}
    \epsilon_0 &= \frac{4}{3(\sigma - 1)^2}, 
    \,\eta_0 = -\frac{4(\sigma - 2)}{3(\sigma - 1)^2}, \\
    \Delta \epsilon &= -\frac{8 \sigma}{9(\sigma - 1)} + \mathcal{O}\left( \gamma \right), 
    \Delta \eta = -\frac{4 \sigma(\sigma + 3)}{9(\sigma - 1)} + \mathcal{O}\left( \gamma \right).
\end{align}
We can calculate $N_e$ by perturbation method:
\begin{align}
    N_e(s_*) &= \int_{s_e}^{s_*} ds\frac{1}{\sqrt{2\epsilon}} \approx\int_{s_e}^{s_*} ds\left(\frac{1}{\sqrt{2\epsilon_0}}-\delta\frac{\Delta\epsilon}{(2\epsilon_0)^{3/2}}\right) \nonumber \\
    &\approx\frac{3}{4}\sigma(s_*)-\frac{3}{4}\ln(\sigma(s_*))+\frac{\delta}{12}\sigma(s_*)^3 .\label{eq:1}
\end{align}

\begin{figure}[t]
    \centering
    \includegraphics[width=0.95\linewidth]{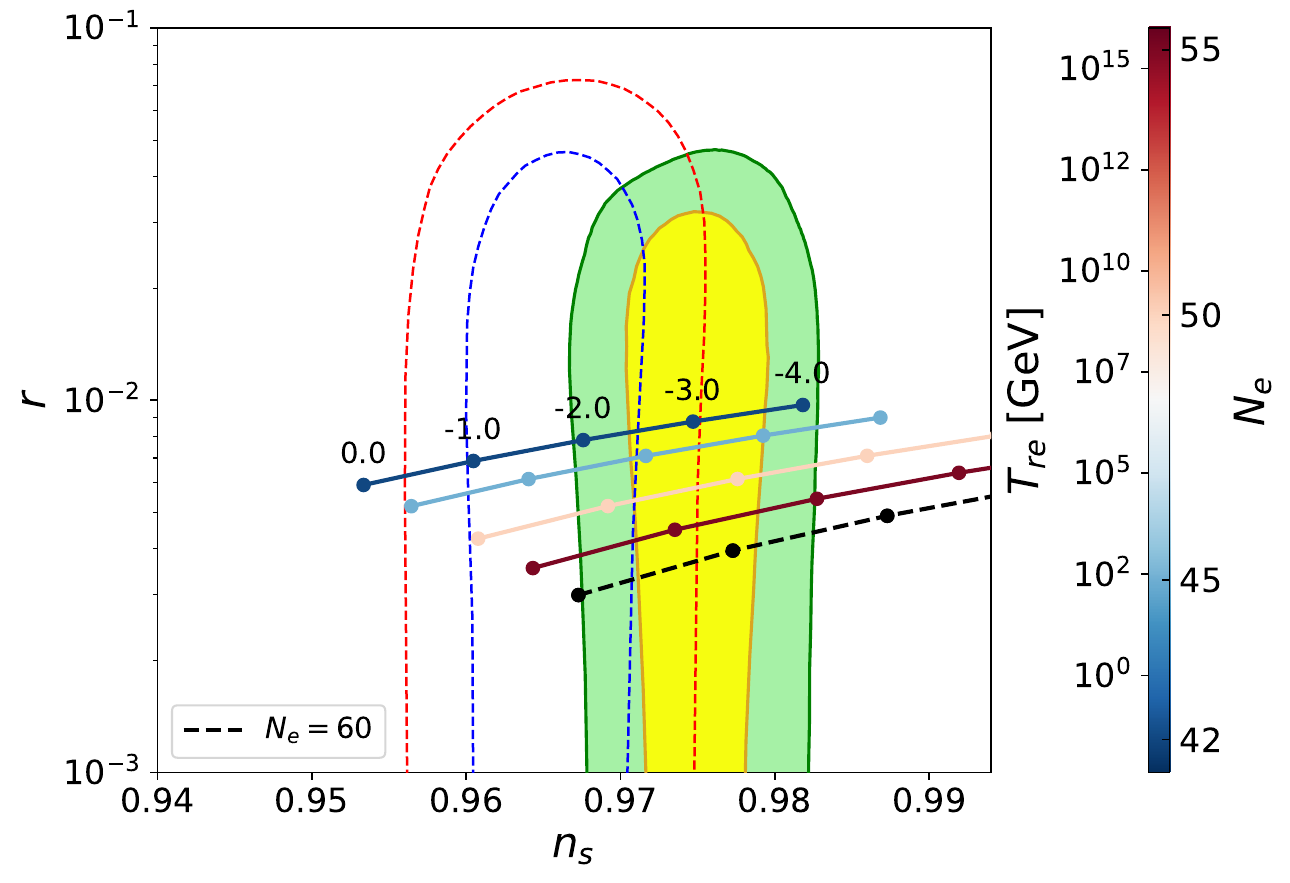}
    \caption{$(n_s, r)$ plane with varying $N_e$ and $\delta$ (in units of $10^{-4}$), under the assumption of matter-dominated reheating ($w_{\rm eff} = 0$). The shaded regions show the $1\sigma$ and $2\sigma$ constraints from P-ACT-LB-BK18 (yellow and green), while the dashed lines denote the $1\sigma$ (blue) and $2\sigma$ (red) bounds from Planck 2018. {The colored solid lines correspond to $N_e=42, 45, 50, 55$, with the color bar indicating the associated reheating temperature $T_{\mathrm {re}}$ and $N_e$. The black dashed line shows the trajectory for $N_e = 60$, which is not included in the color bar.}}
    \label{fig:nsr}
\end{figure}

A recent study~\cite{Drees:2025ngb} suggests that the second term in Eq.~\eqref{eq:1} can affect the results. Therefore, we do not ignore this term but instead treat it as a perturbation.
\begin{align}
\begin{aligned}
    \sigma(s_*)&=\frac{4}{3}N_e(s_*)+\ln\left(\sigma(s_*)\right)-\frac{\delta}{9}\sigma(s_*)^3\\
    &\approx \frac{4}{3}N_e'(s_*)-\delta\frac{64}{243}\left(N_e'(s_*)\right)^3
\end{aligned}
\end{align}
where $N_e'=N_e+\frac{3}{4}\ln\left(\frac{4}{3}N_e\right)$.\\
From this, the spectral index and tensor-to-scalar ratio can be expressed as:
\begin{align}
    n_s &\approx 1 - \frac{2}{N_e'} - \frac{9}{2N_e'^2} - \delta \frac{128}{81} N_e', \\
    r &\approx \frac{12}{N_e'^2} - \delta \frac{256}{27}.
\end{align}
In general, higher order terms deform the potential and lead to sizable corrections to the observables.\footnote{With the $R^4$ contribution, we find the following additional terms:
\begin{align}
\Delta n_s & \approx -\frac{16}{3}N_e'^2
\delta_4,\\
\Delta r& \approx -\frac{64}{3}N_e' 
\delta_4
\end{align}
where $\delta_4\equiv \frac{c_4}{c_2^3}\ll \delta_3=\frac{c_3}{c_2^2}$ encapsulates the degree of deformation due to $R^4$. }

To match the overall amplitude of the scalar power spectrum to the latest observations, we adopt the constraint
$\log\left(10^{10}A_s\right) = 3.060^{+0.011}_{-0.012}$,
as reported by the ACT collaboration \cite{ACT:2025fju}.  Using the slow-roll approximation, $A_s \simeq \frac{V_E}{24\pi^2 \epsilon_0}$ we estimate 
\begin{align}
    \beta\simeq 2.38\times10^9\left(\frac{N_e'}{60}\right)^2.
\end{align}

\begin{figure}[t]
    \centering
    \includegraphics[width=0.9\linewidth]{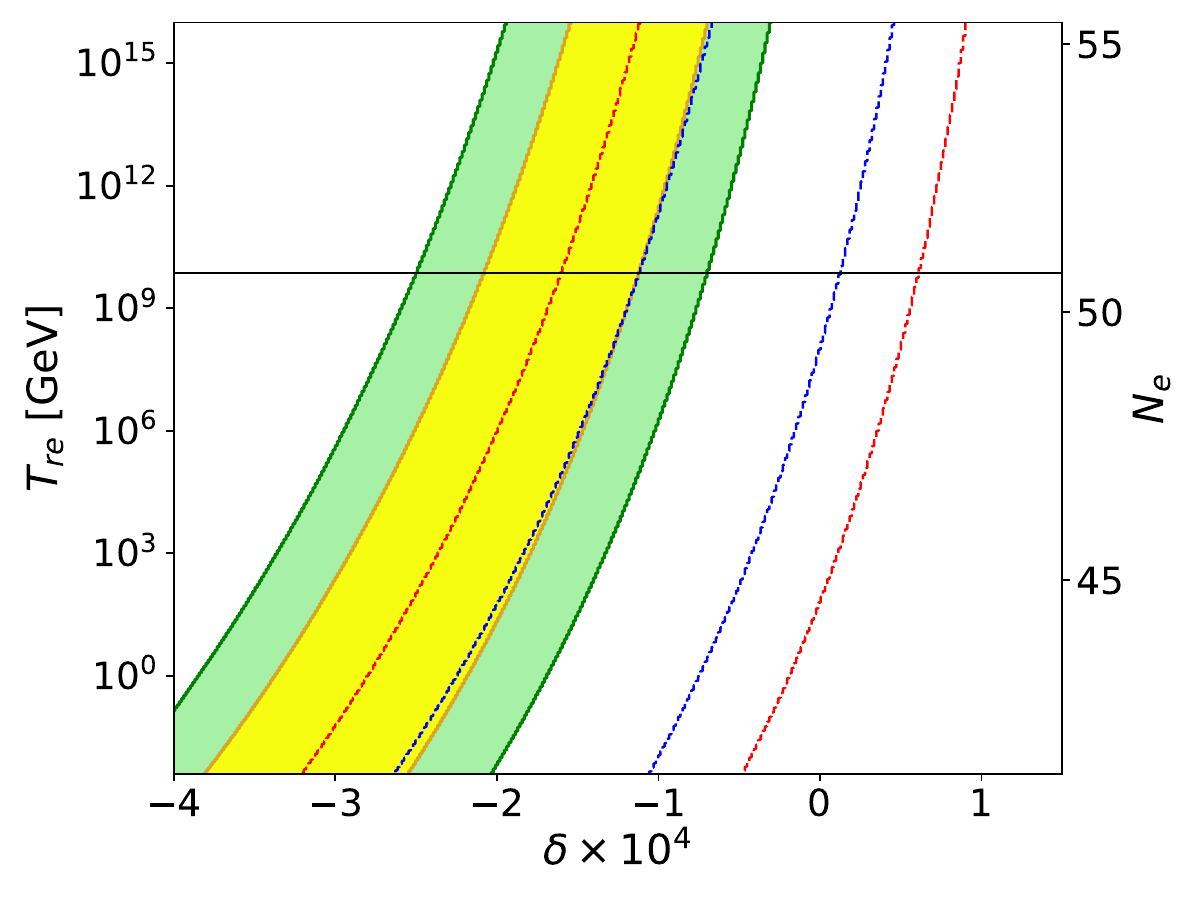}
    \caption{Allowed region in the $(\delta, T)$ plane (with $\delta$ in units of $10^{-4}$) under the assumption of matter-dominated reheating ($w_{\rm eff} = 0$). 
        The shaded regions represent the $1\sigma$ (yellow) and $2\sigma$ (green) constraints from P-ACT-LB-BK18, while the dashed lines denote the $1\sigma$ (blue) and $2\sigma$ (red) bounds from Planck 2018. The horizontal black line indicates the benchmark point corresponding to the perturbative decay scenario with $N_e \simeq 50.6$ and $T_{\mathrm{re}} \simeq 5.1\times10^{9}\,\mathrm{GeV}$.}
    \label{fig:delta_T_w0.0.pdf}
\end{figure}

Fig.~\ref{fig:nsr} presents the theoretical predictions on the $(n_s, r)$ plane for a range of e-folding numbers $N_e$ and perturbation parameters $\delta$. Each trajectory corresponds to a fixed value of $N_e$, color-coded according to the logarithm of the reheating temperature with effective equation of state $w_{\rm eff}=0$. The plot illustrates that $\delta = 0$, which corresponds to the original Starobinsky model, is disfavored at the $2\sigma$ level by the latest P-ACT-LB-BK18 data.\footnote{\dyc{ We note that the region with $N_e\gtrsim60$ is marginally $2\sigma$ compatible~\cite{Drees:2025ngb}, however this will only be viable for $w_{\mathrm{eff}}\gtrsim0.5$. We further discuss this in Section~\ref{sec:Trehdeltarange}.}}  

\section{Reheating: \texorpdfstring{$T_{\mathrm{re}}$}{T re} vs \texorpdfstring{$\delta$}{delta}}
\label{sec:Trehdeltarange}

\begin{figure}[t]
    \centering
    \includegraphics[width=0.9\linewidth]{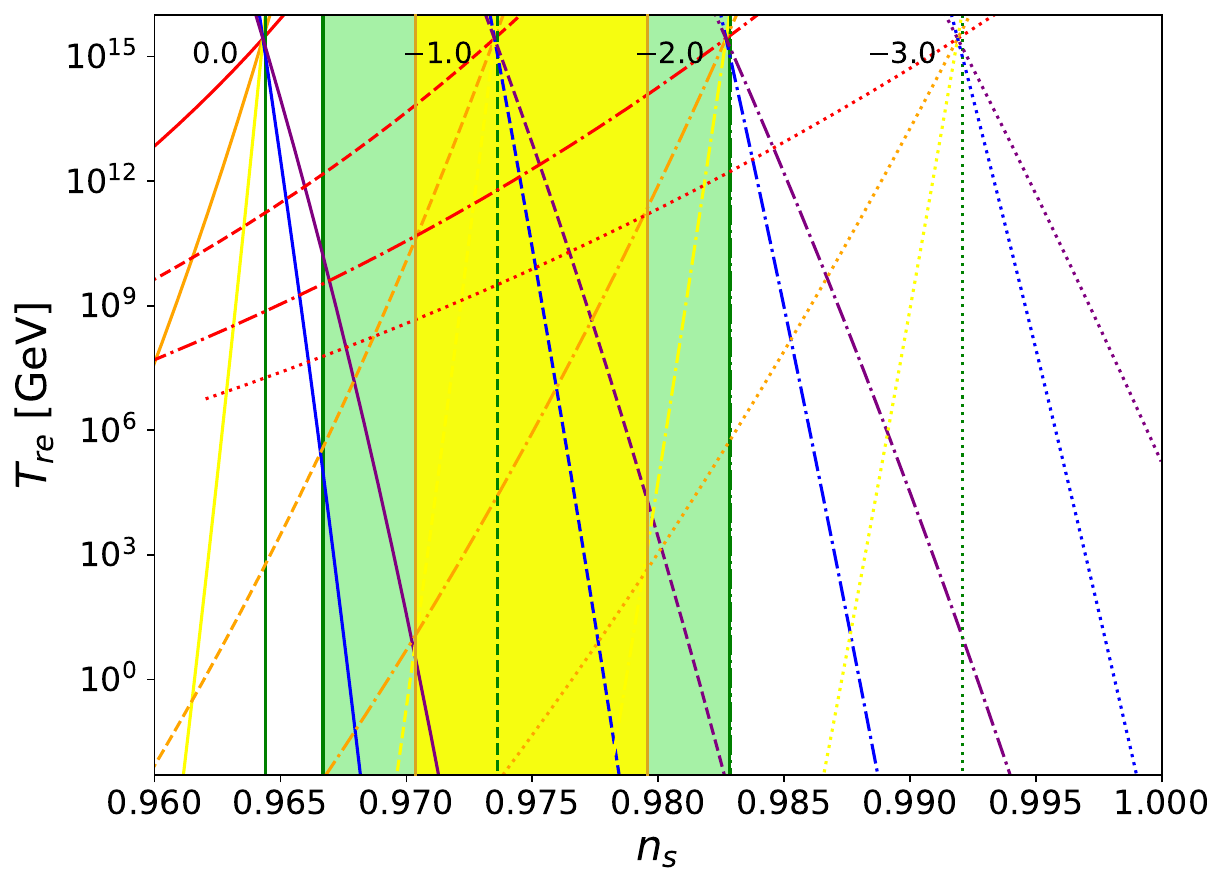}
    \caption{Predictions in the $(n_s, T)$ plane for varying $\delta$ and $w_{\rm eff}$. Each group of five lines corresponds to a fixed value of $\delta $ (in units of $10^{-4}$). Within each group, the lines represent different values of the equation-of-state parameter $w_{\rm eff} = -1/3$ (red), $0$ (orange), $1/5$ (yellow), $1/3$ (green), $3/5$ (blue) and $1$ (purple). The shaded regions correspond to the $1\sigma$ (yellow) and $2\sigma$ (green) bounds on $n_s$ from P-ACT-LB-BK18.}
    \label{fig:nsT}
\end{figure}

In this section, we study the connection between the reheating temperature $T_{\mathrm{re}}$ and the perturbation parameter $\delta$. The number of $e$-folds $N_e$ is related to $T_{\mathrm{re}}$ and the effective reheating equation-of-state parameter $w_{\rm eff}$ through the standard expression \cite{Cook:2015vqa}:


\begin{align}
N_e = 61.4 + \frac{3 w_{\rm eff} - 1}{12(1+w_{\rm eff})} \ln\!\left( \frac{45 V_\mathrm{end}}{\pi^2 g_* T_{\mathrm{re}}^4} \right) - \ln\!\left( \frac{V_\mathrm{end}^{1/4}}{H_*} \right).
\label{eq:Ne}
\end{align}
For a fixed $w_{\rm eff}$, this relation directly links $T_{\mathrm{re}}$ to the field value at horizon exit, and therefore to the inflationary observables. In Fig.~\ref{fig:nsr}, the color bar indicates $T_{\mathrm{re}}$ for $w_{\rm eff}=0$. For a given $w_{\rm eff}$, we can map the observational bounds on $(n_s,r)$ into constraints on the $(\delta, T_{\mathrm{re}})$ plane, as shown in Fig.~\ref{fig:delta_T_w0.0.pdf}.
The reheating temperature is obtained from the perturbative inflaton decay. Within our consideration, this corresponds to the reheating stage being in a matter-dominated phase with an effective equation-of-state parameter $w_{\rm eff}=0$. The interaction Lagrangian is given in terms of the trace of the energy-momentum tensor~\cite{Choi:2019osi}:

\begin{align}
\frac{\mathcal{L}_{\mathrm{int}}}{\sqrt{-g}} & =-\frac{1}{2 f^{\prime}(\phi)} T_\mu^\mu=-\frac{1}{2} e^{-\sqrt{\frac{2}{3} }s} T_\mu^\mu
\end{align}

In this case, the inflaton decay  is dominated by the electroweak sector~\cite{Choi:2019osi}.~\footnote{\dyc{The presence of a non-minimal coupling between the Higgs and the gravity sector $\mathcal{L}/\sqrt{-g} \supset \xi h^2 R/2$ will alter the reheating dynamics with $ \Gamma \simeq {(1+6 \xi)^2 {{m}_s^3}}/{48 \pi}$~\cite{Ema:2019fdd}. We take a conservative approach and focus on $\xi = 0 $ for our analysis.}} 
The corresponding decay rate is
\begin{align}
\Gamma_s \simeq \frac{m_s^3}{48\pi}, \qquad m_s = \frac{1}{\sqrt{3\beta}}.
\end{align}

The reheating temperature is derived as:
\begin{align}
    \begin{aligned}
T_{\mathrm{re}} & =\left(\frac{90}{\pi^2 g_*}\right)^{1 / 4} \Gamma_s^{1/2} \\
&\simeq   (4.36\times 10^{9} ~{\rm GeV})\left(\frac{60}{N_e'}\right)^{3/2},
\end{aligned}
\end{align}
for $g_*=106.75$. 

Combining this with the $T_{\mathrm{re}}$–$N_e$ relation in Eq.~\eqref{eq:Ne}, we obtain the values of $N_e\simeq 50.6$ and $T_{\mathrm{re}}\simeq 5.1\times10^9 \mathrm{GeV}$. Imposing the observed $n_s$ constraint from P-ACT-LB-BK18, the perturbative decay scenario leads to the negative cubic deformation,
\begin{align}
\delta \simeq -1.6\times 10^{-4},
\end{align}
which corresponds to the central value of the region favored by the P-ACT-LB-BK18 $n_s$ constraint, and provides a significantly better fit compared to the $\delta=0$ Starobinsky limit.

We then extend the analysis by scanning over both $\delta$ and $w_{\rm eff}$, imposing only the $n_s$ constraint since $r$ is always within the allowed range for the parameter space of interest. Fig.~\ref{fig:nsT} shows the $(n_s, T_{\mathrm{re}})$ predictions for several $\delta$ values. Each color corresponds to a different $w_{\rm eff}$; the slope and position of the curves indicate how reheating dynamics affect the $n_s$–$T_{\mathrm{re}}$ correlation. 
\dyc{The fixed points in Fig.~\ref{fig:nsT} for each $\delta$ correspond to the case where the entire inflaton energy density has been transferred to radiation.}\footnote{{In Eq.~\eqref{eq:Ne}, the fixed points in Fig.~\ref{fig:nsT} 
correspond to the condition
\begin{align}
    \ln\left(\frac{45 V_{\mathrm{end}}}{\pi^2 g_* T_{\mathrm{re}}^4}\right) = 0
    \;\;\Longrightarrow\;\;
    \frac{\pi^2}{30} g_* T_{\mathrm{re}}^4 = \frac{3}{2} V_{\mathrm{end}}\, .
\end{align}}}.
For $\delta = -1.0 \times 10^{-4}$, all $w_{\rm eff}$ values yield $n_s$ within the $1\sigma$ region, while for $\delta \approx 0$ only relatively large $w_{\rm eff}$ remain viable. For $\delta \lesssim -2.0\times 10^{-4}$, the preferred region shifts toward smaller $w_{\rm eff}$, indicating that stronger cubic deformations favor softer reheating equation of state.

\begin{figure}[t]
    \centering
    \includegraphics[width=0.9\linewidth]{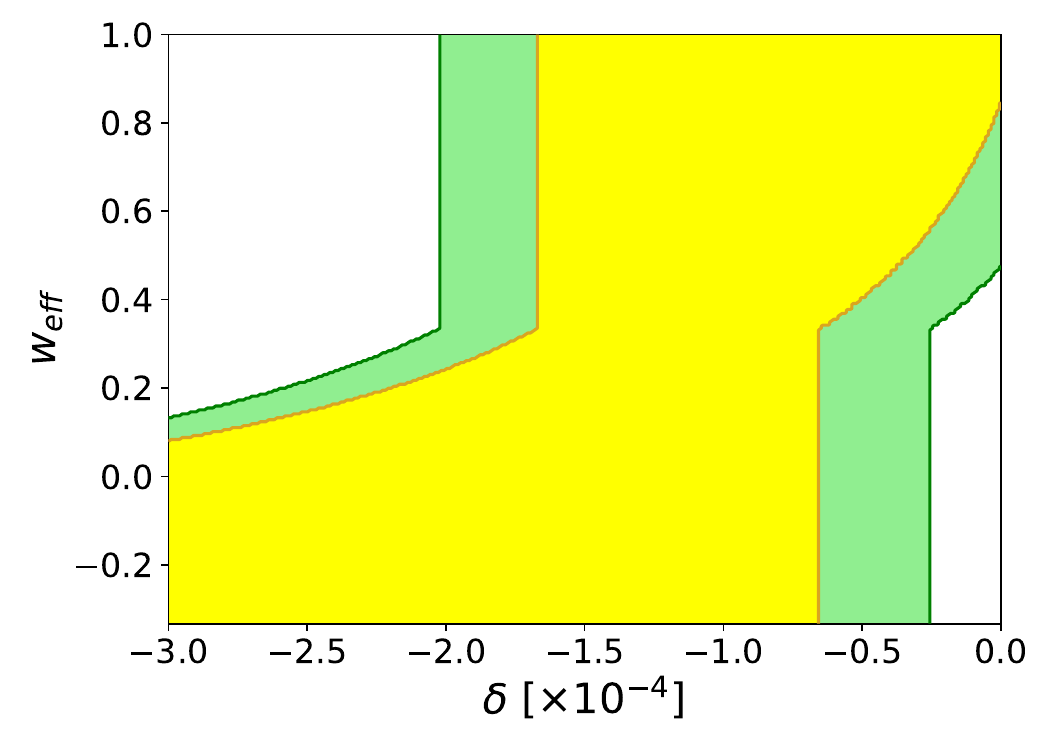}
    \caption{Allowed region in the $(\delta, w_{\rm eff})$ plane, showing the combinations of $\delta$ (in units of $10^{-4}$) and the effective equation-of-state parameter $w_{\rm eff}$ that are consistent with observations. The shaded regions correspond to the $1\sigma$ (yellow) and $2\sigma$ (green) constraints from P-ACT-LB-BK18.}
    \label{fig:delta-w}
\end{figure}

The combined constraints on $\delta$ and $w_{\rm eff}$ are summarized in Fig.~\ref{fig:delta-w}. 
\dyc{In obtaining these bounds we impose a bound $T_{\mathrm{re}}\in[4\times10^{-3},\, 3\times10^{15}]\,\mathrm{GeV}$ so that reheating is completed before BBN ~\cite{deSalas:2015glj} and does not exceed the inflationary energy scale.} 
The $1\sigma$ (yellow) and $2\sigma$ (green) allowed regions form continuous bands that tilt in the $(\delta, w_{\rm eff})$ plane: For $\delta \approx 0$, only relatively large $w_{\rm eff}$ values are allowed, while around $\delta \approx -1\times 10^{-4}$ the full range of $w_{\rm eff}$ is consistent with the $n_s$ constraint. For more negative $\delta$, the allowed range shifts toward smaller $w_{\rm eff}$, with large $w_{\rm eff}$ values being excluded. The vertical boundaries in the $(\delta, w_{\rm eff})$ plane arise when the fixed-$\delta$ trajectories in the $(n_s, T_{\mathrm{re}})$ plane cross the $1\sigma$ or $2\sigma$ observational bands simultaneously for all $w_{\rm eff}$.

\section{Conclusions and Discussions}
\label{sec:conclusions}

We analyzed the $f(R) = R + (\beta/2)R^2 + (\gamma/3)R^3$ inflationary model in light of the most recent P-ACT-LB-BK18 data, treating the cubic term perturbatively via $\delta = \gamma/\beta^2 \ll 1$. Analytic expressions for $n_s$ and $r$ including $\mathcal{O}(\delta)$ corrections were obtained, and updated observational constraints were used to map out the allowed $(\delta, N_e, T_{\mathrm{re}}, w_{\rm eff})$ parameter space.

Our results indicate that the original Starobinsky model ($\delta = 0$) is mildly disfavored at the $2\sigma$ level, while small negative values of $\delta$ improve compatibility with the observed spectral tilt. In particular, for perturbative reheating with $w_{\rm eff} = 0$, we find that $T_{\mathrm{re}} \simeq 5.1\times 10^{9}$ GeV and $\delta \simeq -1.6\times 10^{-4}$ yields a significantly better fit to the ACT data. More generally, the $(\delta, w_{\rm eff})$ parameter space exhibits characteristic vertical boundaries, arising when fixed-$\delta$ trajectories in the $(n_s, T_{\mathrm{re}})$ plane intersect the $1\sigma$ or $2\sigma$ observational bands across all $w_{\rm eff}$ values (see Fig.~\ref{fig:delta-w}). Notably, $\delta \sim -10^{-4}$ permits the full range of $w_{\rm eff}$, whereas $\delta \approx 0$ or more negative values impose strong restrictions on reheating dynamics.

Our results demonstrate that even small cubic deformations of the Starobinsky model can alleviate tensions with current CMB data while simultaneously placing nontrivial constraints on the reheating phase. We emphasize that our analysis remains strictly within the $f(R)$ framework based on the Ricci scalar. Nonetheless, extensions involving higher-curvature invariants, such as Gauss–Bonnet terms (e.g.~\cite{Koh:2023zgn}), remain interesting alternatives for future investigation.

\mbox{}

\acknowledgements
This work was supported by the National Research Foundation of Korea(NRF) grant funded by the Korea government (MSIT) (RS-2023-00283129), (RS-2024-00340153) and Yonsei internal grant for Mega-science (2023-22-048). 
We thank Hyun Min Lee for the earlier collaboration.

\bibliographystyle{utphys.bst}
\bibliography{ref}

\end{document}